\documentclass[letterpaper]{article}

\usepackage{authblk}
\usepackage{amsmath}
\usepackage{amssymb}
\usepackage{bbm}
\usepackage{bm}
\usepackage[labelfont=bf]{caption}
\usepackage{enumerate}
\usepackage{graphicx,psfrag,epsf}
\usepackage{lscape}
\usepackage{multicol} 
\usepackage{multirow}
\usepackage[square,sort,comma,numbers]{natbib}
\usepackage{setspace}
\usepackage{url}

\usepackage{epstopdf}

\pdfoutput=1

\title{\bf Outcome identification in electronic health records using predictions from an enriched Dirichlet process mixture}
  
\author[1]{ Bret Zeldow}
\author[2]{James Flory}
\author[3]{Alisa Stephens-Shields}
\author[4]{Marsha Raebel}
\author[3]{Jason Roy}

\affil[1]{\footnotesize Department of Health Care Policy, Harvard Medical School, Boston, MA, USA}
\affil[2]{\footnotesize Department of Medicine, Weill Cornell Medical College, New York, NY, USA}
\affil[3]{\footnotesize Department of Biostatistics, University of Pennsylvania, Philadelphia, PA, USA}
\affil[4]{\footnotesize Institute for Health Research, Kaiser Permanente Colorado, Aurora, CO, USA}

\date{}

\begin {document}
  \maketitle

\begin{abstract}
We propose a novel semiparametric model for the joint distribution of a continuous longitudinal outcome and the baseline covariates using an enriched Dirichlet process (EDP) prior. This joint model decomposes into a linear mixed model for the outcome given the covariates and marginals for the covariates. The nonparametric EDP prior is placed on the regression and spline coefficients, the error variance, and the parameters governing the predictor space. We predict the outcome at unobserved time points for subjects with data at other time points as well as for new subjects with only baseline covariates. We find improved prediction over mixed models with Dirichlet process (DP) priors when there are a large number of covariates. Our method is demonstrated with electronic health records consisting of initiators of second generation antipsychotic medications, which are known to increase the risk of diabetes. We use our model to predict laboratory values indicative of diabetes for each individual and assess incidence of suspected diabetes from the predicted dataset. Our model also serves as a functional clustering algorithm in which subjects are clustered into groups with similar longitudinal trajectories of the outcome over time.
\end{abstract}

\noindent%
\vfill

\newpage
\section{Introduction}

Electronic health records (EHR), now a critical component of health care, make a large quantity of data available for researchers. Challenges in using EHR for statistical analyses, however, are well-documented \cite{national2017refining}. 
The focus of this paper is on the challenge of outcome identification. Many diseases can be identified in the data from diagnostic codes. However, this is unlikely to fully capture outcomes. EHR data often contain longitudinal measures from laboratory tests (labs) which can be used for the diagnosis of diseases and for disease monitoring. In practice, labs are sometime used to identify additional outcomes (beyond those identified from diagnostic codes). For instance, subjects at risk for diabetes can have fasting glucose labs monitored over time, which can be instrumental in diagnosing the disease \cite{american2014diagnosis}. From a statistical perspective, one challenge is that labs may be abundant for some subjects and sparse or missing for others. Unlike in planned observational studies with primary data collection, labs are not necessarily observed at ideal times. Correspondingly, it may be helpful to model these labs and to use this model to make predictions at time points of interest for EHR containing missing or sparse data. To this end, we propose a flexible joint model for the distribution of a continuous longitudinal outcome (lab values) and baseline covariates. The parameters from the joint model are all given a Dirichlet process (DP) prior with the enrichment proposed in \citet{wade2011enriched}. Our model provides a flexible framework for prediction as well as serving as a functional clustering algorithm in which one does not specify the number of clusters \textit{a priori}.


The Dirichlet process (DP) mixture is a popular Bayesian nonparametric (BNP) model \cite{ferguson1973bayesian, ferguson1983bayesian, escobar1995bayesian} found in many applications, including topic modeling \cite{teh2004sharing}, survival analysis \cite{hanson2004bayesian}, regression \cite{hannah2011dirichlet}, classification \cite{cruz2007semiparametric}, and causal inference \cite{roy2017bayesian}. Consider the regression setting of \citet{shahbaba2009nonlinear} and \citet{hannah2011dirichlet}, where there is an outcome $Y$ which we would like to regress on covariates $X$. In a Bayesian generalized linear model (GLM) setup \cite{mccullagh1984generalized}, the predictors $X$ are restricted to be a linear combination of the unknown regression parameters. Because of this, GLMs are not appropriate to model nonlinear response curves when the regression coefficients are given normally distributed priors \cite{gelman2014bayesian}. In contrast, placing a DP prior on the regression coefficients (DP-GLM) instead of a parametric prior allows for nonlinearities despite the underlying GLM framework, and this flexibility can often be achieved with only modest additional computational burden. The power of the DP prior stems in part from its partitioning properties \cite{muller2015bayesian}, where it clusters observations and fits local regressions among subjects with similar relationships between covariates and the outcome \cite{hannah2011dirichlet}.

\citet{wade2011enriched} showed that with a high number of covariates $X$, the likelihood contribution of $X$ can dominate the posterior of the partition so that clusters form based more on similarity of covariates than on regression parameters. This leads to a high number of clusters with few observations per cluster and can result in poor predictive performance that can be improved by using an enriched DP (EDP) mixture instead of a DP mixture \cite{wade2014improving}. The EDP mixture allows for nested clustering, where one can have clusters based solely on the regression coefficients governing $Y$ on $X$ and within those, nested clusters based on similarity in the covariate space. The benefits of the EDP mixture were demonstrated in simulation and in a real data analysis \cite{wade2014improving}.

In this paper, we extend the EDP mixture model to longitudinal settings with a continuous outcome. Some alternatives to our EDP approach to longitudinal data have been proposed in the literature. \citet{muller1997bayesian} modeled blood concentrations in a pharmacokinetic study using DP mixtures with a DP prior on the covariate parameters and the regression coefficients. \citet{li2010bayesian} developed a flexible semiparametric mixed model with smoothing splines and a DP prior on the random effects with a uniform shrinkage prior for its hyperparameters. \citet{das2013bayesian} fit a bivariate longitudinal model for sparse data with penalized splines for the effect of time and DP priors on the random effects. \citet{quintana2016bayesian} developed a longitudinal model with random effects and a Gaussian process with DP mixtures on covariance parameters of the Gaussian process. This allows for flexible modeling of the correlation structure. \citet{bigelow2009bayesian} fit a joint model for a binary outcome and functional predictor where the functional predictor was modeled with cubic B-splines whose basis coefficients were given a DP prior. \citet{scarpa2014enriched} developed an enriched (unrelated to the enriched DP) stick-breaking process which incorporated curve features to better fit functional data. 


Our model is unique in that the regression parameters and the parameters for the covariates are given an EDP prior rather than the usual DP prior. As a result, the partitions are not dominated by the covariates as may otherwise happen. Along with improved prediction over DP priors, our model serves as a functional clustering algorithm in which subjects with similar trajectories over time are likely to be part of the same cluster. This aspect also benefits from the EDP prior as functions cluster separately on the regression parameters and covariates. Functional clustering can illuminate distinct patterns among different groups of subjects. A review of functional clustering can be found in \citet{jacques2014functional}. Notably, frequentist and parametric Bayesian methods often require prior specification of the number of clusters, often chosen through model fit statistics. Our EDP model requires no such specification; new clusters may form and existing clusters may vanish throughout the Markov Chain Monte Carlo (MCMC) algorithm.

Our motivating example is a study of individuals who newly initiate a second-generation antipsychotic (SGA). SGAs are known to increase incidence of diabetes \cite{newcomer2005second, de2012metabolic}. A previous analysis explored the value of incorporating elevated laboratory test results as part of the definition of the outcome of incident diabetes, defined by diagnosis codes and dispensing claims of antidiabetics \cite{flory2017missing}. However, many subjects had no recorded lab values or had them measured outside the narrow study window. In this paper, we demonstrate our model by regressing each of three lab values indicative of diabetes (hemoglobin A1c, fasting glucose, and random glucose) on baseline covariates and time. Throughout the MCMC algorithm, values are predicted for each subject at the end of the individual's follow-up, either at one year post SGA initiation or earlier if censored prior to that time. We then combine each set of predictions with the observed data so that each draw can be thought of as an imputed lab value. We then calculate the incidence of diabetes using a multiple imputation procedure \cite{rubin2004multiple}. Lastly, we demonstrate how our model can be used for functional clustering by examining posterior clustering patterns resulting from the model.

The rest of the paper is organized as follows. In section 2, we write out the details of our model and describe key components. In section 3, we discuss computations and making predictions from our model. In section 4, we test our method on simulated datasets. In section 5, we apply our method to the SGA dataset. We discuss the paper in section 6 including limitations and future directions.

\section{Model}
To motivate our model, first consider a hypothetical planned observational study, where the outcome of interest is diabetes status one year following initiation of an SGA. In that hypothetical study, we would collect laboratory data, such as hemoglobin A1c (HbA1c) at the end of the study. We might then classify people as having the outcome if their HbA1c value was $\geq$ 6.5\%.

\begin{figure}
\centering
\includegraphics[width=0.7\textwidth]{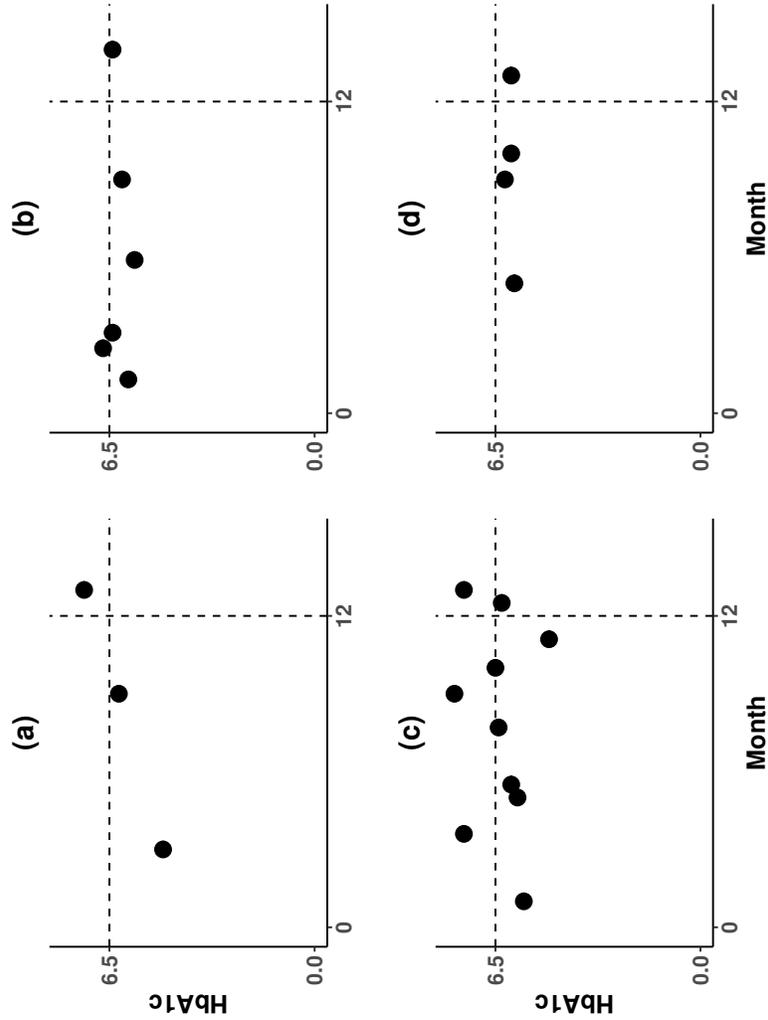}
\caption{Hypothetical example of data from electronic health records. The four panels represent the hemoglobin A1c (HbA1c) values for four subjects over a 12+ month period. The vertical dashed line at month 12 indicates the time point of interest. The horizontal dashed line represents the critical value (6.5\%) of HbA1c above which or equal to indicates diabetes. The cross marks indicate the observed values for each subject. None of the four subjects have values taken precisely at month 12 so interpolation is necessary. Panel (a) shows a subject who has a rising trajectory but doesn't cross the threshold until after month 12. Panel (b) shows a subject who crosses the threshold once prior to month 12 but is stable below the threshold for many other observations. Panel (c) shows a subject with highly variable data around the threshold before and after month 12. Panel (d) shows a subject below the threshold for all observed data.}
\label{fig:ehr}
\vspace{-14.01336pt}
\end{figure}

Now consider a study with the same goals, but using EHR data. Figure~\ref{fig:ehr} shows four hypothetical subjects with longitudinal measurements of HbA1c over a period of about 15 months. We are interested in determining whether HbA1c levels are $\geq$ 6.5\% at month 12. However, none of the four subjects have data collected precisely at month 12, so we need to interpolate from observed data to classify them as elevated or not at month 12. How we classify them is dependent on the algorithm used. Naive algorithms might include basing classification on the value closest to month 12, on the value closest to month 12 that is prior to month 12, or on the maximum value prior to month 12. For instance, it is clear that subject (a) can be classified as either elevated or not depending on the algorithm implemented. Subject (b) has many observations but only one is above the critical threshold and the overall trend suggests their value at month 12 would not elevated. The data for subject (c) has highly variable data and it is uncertain what their month 12 value would be. All subjects have varying degrees of uncertainty in their classifications. These naive classification methods do not use all of the data and do not account for uncertainty in the prediction/imputation.

Our BNP model, described below, was designed to impute outcomes at any time or times of interest, while fully utilizing all of the data (covariates and labs over time). It uses all available data and predicts the outcome at unobserved time points periodically throughout the MCMC algorithm. Thus, for each subject we estimate the distribution of the outcome at the time point of interest rather than just a single prediction.

\subsection{Notation}

Let $y_{ij}$ denote the $j^{\text{th}}$ occurrence ($1 \leq j \leq n_i$) of a continuous outcome for subject $i$, $i \in [1, \dots, n]$, observed at time $t_{ij}$. Let $\mathbf{y}$ denote the vector of outcomes for all subjects and $\mathbf{y}_i$ denote the vector of outcomes for the $i^{\text{th}}$ subject. Let $\mathbf{t}_i$ denote the vector of time points at which $\mathbf{y}_i$ were recorded so that both $\mathbf{t}_i$ and $\mathbf{y}_i$ are length $n_i$. The covariates for subject $i$ are measured at baseline and denoted by the $p$-dimensional vector $\mathbf{x}_i$. Without loss of generality, let the first $p_1$ values $\mathbf{x}_i$ be binary and the remaining $p_2$ be continuous with $p = p_1 + p_2$. Let $n$ denote the total number of subjects and $N$ denote the total number of observations, accounting for multiple observations per subject. 

We model the distribution the outcome $y_{ij}$ as a function of covariates $\mathbf{x}_i$ and time $t_{ij}$ jointly with the marginal distributions of $\mathbf{x}_i$. To allow for nonlinearities across time, we use splines with $k$ prespecified knots at $(q_1, \dots, q_k)$ with $q_1 \leq \dots \leq q_k$. \citet{bigelow2009bayesian} considered B-splines \cite{hastie1990generalized} and \citet{li2010bayesian} used P-splines. We opt for penalized, thin plate splines which have good mixing properties in Bayesian analysis \cite{crainiceanu2005bayesian}. The choice of penalized splines also allows us to choose a large number of knots, reducing the dependency of the model fit on the selection of knot locations. However, any number of basis expansions are possible, including wavelets \cite{ray2006functional}. For thin plate splines, let $\mathbf{Z}$ denote the $N$ by $k$ matrix with each row corresponding to the basis functions evaluated at each observed time point $t$. The matrix $\mathbf{Z}$ is calculated as $\mathbf{Z} = \mathbf{Z}_k \Omega_k^{-1/2}$, where the rows of $\mathbf{Z}_k$ are equal to $\left\{|t_{ij} - q_1|^3, \dots, |t_{ij} - q_k|^3\right\}$ and the penalty matrix $\Omega_k$ is a $k \times k$ matrix where the $(l, m)$th entry is $|q_l - q_m|^3$ \cite{crainiceanu2005bayesian}. The penalty matrix prevents overfitting by penalizing the coefficients of $Z_k$. Each subject $i$ in the sample contains a $n_i$ by $k$ submatrix $\mathbf{z}_i$ of $\mathbf{Z}$ which corresponds to the basis functions evaluated at each $t_{ij}$. 

We fit the model
\begin{align}
\mathbf{y_i} | \mathbf{x_i}, \mathbf{t_i}, \bm{\beta_i}, \bm{\eta_i}, u_i, \sigma^2_i & \sim \text{N}(\mathbf{x_i^*}\bm{\beta_i} + \mathbf{z_i} \bm{\eta_i} + u_i, \sigma^2_{i} \mathbf{I}), \label{eq:y}\\
x_{ij} | \bm{\psi_{i}} & \sim \text{N}(\mu_{ij},\sigma^2_{\mu,ij}) \text{      (for continuous covariates)}; \label{eq:xcont}\\
x_{ij} | \bm{\psi_{i}} & \sim \text{Bernoulli}(p_{ij})  \text{    (for binary covariates)};\label{eq:xbin}\\
u_i &\sim \text{N}(0, \sigma^2_u); \nonumber \\
(\bm{\theta_i}, \bm{\psi_i}) | P & \sim P; \nonumber \\
P & \sim \text{EDP}(\alpha_{\theta}, \alpha_{\psi}, P_0); \nonumber \\
\sigma^2_u,  \alpha_{\theta}, \alpha_{\psi} & \sim \text{Inv-Ga}(a_u, b_u) \times \text{Ga}(a_{\theta}, b_{\theta}) \times \text{Ga}(a_{\psi}, b_{\psi}); \nonumber 
\end{align}
where $\bm{\theta_i} = (\bm{\beta_i},\sigma^2_{\beta, i},\bm{\eta_i},\sigma^2_{\eta, i},\sigma^2_i)$ are the regression parameters. The notation $EDP(\alpha_{\theta}, \alpha_{\psi}, P_0)$ means that $P_{\theta} \sim DP(\alpha_{\theta}, P_{0 \theta})$ and $P_{\psi|\theta} \sim DP(\alpha_{\psi}, P_{0\psi|\theta})$, where $\alpha_{\theta}$ and $\alpha_{\psi}$ are positive valued parameters and $P_0 = P_{0 \theta} \times P_{0\psi|\theta}$ is the base distribution with parameters $\bm{\psi}$ and $\bm{\theta}$ independent. Here,
\[
P_{0 \theta} \sim \underbrace{\text{Inv-Ga}(a_{\beta}, b_{\beta})}_{\sigma^2_{\beta}} \ \times \
                            \underbrace{\text{N}(\bm{\beta_0},\sigma^2_{\beta,i} \mathbf{I})}_{\beta} \ \times \
                            \underbrace{\text{Inv-Ga}(a_{\eta}, b_{\eta})}_{\sigma^2_{\eta}} \ \times  \
                            \underbrace{\text{N}(0, \sigma^2_{\eta,i} \mathbf{I})}_{\eta}  \ \times \
                            \underbrace{\text{Inv-Ga}(a_{y}, b_{y})}_{\sigma^2};
\] and
\[
P_{0\psi|\theta} \sim \prod_{i = 1}^{p_1} \text{Beta}(a_x, b_x) \times \prod_{i = p + 1}^{p_1 + p_2} \text{scaled Inv-}\chi^2(\nu_0, \tau_0^2) \times \text{N}(\mu_0, \tau^2/c),
\]
where the first product is among binary covariates followed by a product over the continuous covariates. The notation $\mathbf{x}^*_i$ indicates the vector $\mathbf{x}_i$ with time $t_{ij}$ possibly added, as would be the case if splines were omitted.

We assume that continuous variables are (locally) normally distributed and that binary predictors are Bernoulli. Other distributions can be used, but these distributions are convenient for their conjugacy properties. The parameter $\bm{\psi}_{i, j}$ corresponds to the two dimensional parameter with mean $\mu_{ij}$ and variance $\sigma^2_{\mu, ij}$ if the $j^{th}$ covariate is continuous or the one dimensional probability parameter $p_{ij}$ if the $j^{th}$ covariate is binary. 
Integrating out the subject specific parameters $\bm{\psi}_i$ and $\bm{\theta}_i$ as in \citet{wade2014improving}, our model can be thought as a countable mixture of linear mixed models where each subject is assigned to one of the mixture components. 

We do not posit any \textit{a priori} relationship between time and the outcome. In some applications where the overall trend may be known (for example, the amount of medication in blood may decrease over time after a drug is administered in a pharmacokinetic study), we may posit a model for equation (\ref{eq:y}) incorporating such knowledge, as in \citet{muller1997bayesian} which assumed a piecewise linear structure.

\subsection{Clustering}

A consequence of using the EDP prior on the regression coefficients is that subjects cluster based on their regression parameters $\bm{\theta}_i$ (that is, for some $i \neq j$, $\bm{\theta}_i = \bm{\theta}_j$), and within these clusters, will form sub-clusters based on their covariate parameters $\bm{\psi}_i$. Since $\bm{\theta}_i$ includes $\bm{\eta}_i$, the coefficients on the spline basis functions for time, subjects with similar trajectories of their outcomes over time will likely be assigned the same cluster. However, subjects are also clustered by the parameter $\sigma^2_i$, which governs variability of outcomes. Thus, it is possible to have clusters with small variability that follow a precise trajectory over time, and it is possible to have clusters whose large variability defines the cluster, or some combination of the two. The total number of clusters depend on the data and the parameters $\alpha_{\theta}$ and $\alpha_{\psi}$, where values closer to 0 indicate fewer clusters.

For this paper, we use the term $\theta$-cluster to indicate clusters based on the parameters $\bm{\theta}$. A $\psi$-cluster denotes a cluster nested within a $\theta$-cluster and indicates closeness in the covariate space governed by covariate parameters $\bm{\psi}$. The $\psi$-clusters are only meaningful with respect to the $\theta$-cluster in which it is nested.

While an advantage of the BNP approach is not having to select the number of clusters, this creates added difficulty in summarizing the clusters. 
We use the strategy employed in \citet{medvedovic2002bayesian}, which employed a distance metric based off of empirical pairwise probabilities of subjects being in the same cluster. To do this, we create a $n \times n$ matrix where each element indicates the number of times two corresponding subjects were in the same $\theta$-cluster over all post burn-in MCMC iterations. From the rows of this matrix, we compute a distance matrix using the supremum norm. We then use Ward's hierarchical agglomerative clustering method implemented by \citet{murtagh2014ward}. This last step requires choosing a number of clusters, which we choose from the median of the posterior distribution on the number of $\theta$-clusters. R code for this calculation is provided in the appendix.

\section{Computations}

Draws from the posterior distribution of all parameters are obtained through Gibbs sampling. We use an extension of algorithm 8 by Neal (2000) \cite{neal2000markov} accommodating the nested partitioning of the EDP \cite{wade2014improving} and repeated measurements. Algorithm 8 involves generating $m$ sets of auxiliary parameters corresponding to $m$ clusters that currently have no members. Broadly, at each iteration we alternate between updating cluster membership for each subject, and then within each cluster we update the parameters $(\bm{\theta}_i, \bm{\psi}_i)$. Let $s_i = (s_{i,y}, s_{i,x})$ denote the cluster membership for the $i^{th}$ subject, where $s_{i,y}$ denotes the $\theta$-cluster corresponding to $\bm{\theta}_i$ and $s_{i,x}$ denotes the $\psi$-cluster nested within $s_{i,y}$ corresponding to $\bm{\psi}_i$. Let $\bm{\theta}^*_k$ denote the value of $\bm{\theta}$ corresponding to the $k^{th}$ unique value of $s_{i,y}$. Similarly, let $\bm{\psi}^*_{j|k}$ denote the value of $\bm{\psi}$ corresponding to the $j^{th}$ unique value of $s_{i,x}$ within the $k^{th}$ unique value of $s_{i,y}.$ Note that if $s_{i, y} = s_{j,y}$, then $\bm{\theta}_i = \bm{\theta}_j$. Furthermore, let $n_k^{\theta}$ denote the number of subjects in the $k^{th}$ unique cluster of $s_{i,y}$ and $n_{j|k}^{\psi}$ denote the number of subjects in the $j^{th}$ unique cluster of $s_{i,x}$ nested within the $k^{th}$ unique value of $s_{i,y}.$ The notation $n_k^{-i, \theta} $ and $n^{-i, \psi}_{j|k}$ denote the size of the clusters with the $i^{th}$ subject removed. Recall that the similar notation with no superscript, $n_i$, refers to the number of observations for the $i^{th}$ individual.

The first step of our algorithm updates the value of $s_i$ for every individual. First, remove individual $i$ from their current cluster. The probability that an individual is in any given cluster depends on the current values of $\alpha_{\theta}$ and $\alpha_{\psi}$, the number of subjects within that cluster, the values of $\bm{\theta}^*$ and $\bm{\psi}^*$ as well as the observed data. In choosing clusters, there are three possibilities: subjects can be assigned to an existing $\psi$-cluster within an existing $\theta$-cluster, a new $\psi$-cluster within an existing $\theta$-cluster, or a new $\theta$-cluster and a new $\psi$-cluster. An individual is assigned to an existing cluster $(k, j)$ with probability proportional to: 
\[
\frac{ n_k^{-i, \theta} n^{-i, \psi}_{j|k} } {(n_k^{-i, \theta} + \alpha_{\psi}) (\alpha_{\theta} + n - 1)}  \times \prod^{n_i}_{v=1} f_y(y_{i,v}; \bm{x}_i, \bm{\theta}^*_k) \times \prod_{l=1}^p f_{x, l}(x_{i, l}; \bm{\psi}^*_{j|k}).
\]
An individual is assigned to  a new $\psi$-cluster within the $k^{th}$ existing $\theta$-cluster with probability proportional to:
\[
\frac{ n_k^{-i, \theta} \alpha_{\psi} / m} {(n_k^{-i, \theta} + \alpha_{\psi}) (\alpha_{\theta} + n - 1)}  \times \prod^{n_i}_{v=1} f_y(y_{i,v}; \bm{x}_i, \bm{\theta}^*_k) \times \prod_{l=1}^p f_{x,l}(x_{i, l}; \bm{\psi}^*_0).
\]
 An individual is assigned a new $\theta$-cluster and a new $\psi$-cluster with probability proportional to:
 \[
\frac{  \alpha_{\theta} / m} { \alpha_{\theta} + n - 1}  \times \prod^{n_i}_{v=1} f_y(y_{i,v}; \bm{x}_i, \bm{\theta}^*_0) \times \prod_{l=1}^p f_{x,l}(x_{i, l}; \bm{\psi}^*_0).
\]
These probabilities are then normalized to sum to 1.

The notation $\bm{\psi}^*_0$ and $\bm{\theta}^*_0$ refers to parameters from a cluster that currently has no members (also called auxiliary parameters, see \cite{neal2000markov}). They are generated randomly from the prior base distributions $P_{0\psi|\theta}$ and $P_{0\theta}$ for $\bm{\psi}$ and $\bm{\theta}$. The notation $f_{x, l} (\cdot; \bm{\psi})$ corresponds to the normal density in equation (\ref{eq:xcont}) or the binomial density in equation (\ref{eq:xbin}) for continuous and binary, respectively, and $f_y(\cdot; \bm{x}_i, \bm{\theta})$ corresponds to the normal density from equation (\ref{eq:y}) evaluated with parameters $\bm{\theta}$. Once we calculate these probabilities, we draw cluster membership using a random multinomial distribution. This is done separately for each individual in the cohort.

Once cluster memberships for all individuals have been updated, the within cluster parameters $\bm{\theta}^*$ and $\bm{\psi}^*$ are updated. To update the regression parameters $\bm{\theta}^*_k$ for the $k^{th}$ cluster, we consider only individuals with $s_{i,y} = k$. First, we update the regression variance $\sigma^{2*}_k$ using a conjugate draw from an inverse gamma distribution and then update regression parameters $\bm{\beta}_k^*$ for covariates $\mathbf{x}$ from a draw with a multivariate normal distribution. Next, update the variance for the spline effects $\sigma^{2*}_{b, k}$ from a random draw from an inverse gamma distribution. Lastly, we update the coefficients $\bm{\eta}_k^*$ for the spline effects from a draw from a multivariate normal distribution. In essence, within each cluster we are fitting separate Bayesian mixed effects models and updating parameters accordingly \cite{zeger1991generalized}. Full posterior distributions for updating $\bm{\theta}^*$ are in the appendix.

Next, we update covariate parameters $\bm{\psi}^*$. To update $\bm{\psi}^*_{j|k}$, we take subjects with $s_i = (k, j)$. If the $l^{th}$ covariate is binary, then the distribution of $x_l$ is assumed Bernoulli and the parameter $\psi_l$ is updated from a Beta distribution with parameters $a_n = \sum_{s = (k, j)} x_{i, l} + a_x$ and $b_n = n_{j|k} - \sum_{s = (k, j)} x_{i, l} + b_x$. If the $l^{th}$ covariate is continuous then the distribution of $x_l$ is normal and the parameters $\psi_l = (\sigma^2_l, \mu_l)$ are updated from conjugate inverse-$\chi^2$ and normal distributions, available in the appendix.

It remains to update the random intercepts $u_i$, the variance $\sigma^2_u$, $\alpha_{\psi}$, and $\alpha_{\theta}$. The new random intercepts $u_i$ are calculated after taking the residuals from the current fit given covariates $\mathbf{x}_i$ and the rest of the current parameter values. The variance $\sigma^2_u$ is updated through a random draw from an inverse gamma distribution with shape $a_u + n$ and rate $b_u +  \frac{\mathbf{u}^{\intercal} \mathbf{u}}{2}$. Finally, we update $\alpha_{\psi}$ and $\alpha_{\theta}$. $\alpha_{\theta}$ is updated by generating a random value from a mixture of two gamma posteriors as in \citet{escobar1995bayesian}. $\alpha_{\psi}$ is updated through a Metropolis-Hastings step. The updates for these $\alpha$ parameters are equivalent to those in \citet{roy2017bayesian} who also employ an EDP mixture model. Consult the appendix for expanded details of the MCMC algorithm.

\subsection{Predictions}

Predicting values for subjects who have observed data (that is, data at time points other than the time point of interest) is straightforward. At every iteration where we seek to make a prediction, each subject is assigned a cluster $s_{i,y}$ with corresponding $\bm{\theta}_i$. From this, we can predict from a single draw from a normal distribution given $\mathbf{x_i^*}, \bm{\beta_i}, \mathbf{z_i}, \bm{\eta_i}, u_i, \sigma^2_i$ with mean $\mathbf{x_i^*}\bm{\beta_i} + \mathbf{z_i} \bm{\eta_i} + u_i$ and variance $\sigma^2$.

For subjects missing outcome data, we must make predictions from their covariates $\mathbf{x_i}$. These subjects may be part of the $k^{th}$ existing cluster with parameters $\bm{\theta}^*_k$ or may be in an entirely new cluster. If they are part of the $k^{th}$ existing $\theta$-cluster, we use the current values from the corresponding parameters for that cluster (i.e., $\bm{\theta}^*_k$) and draw the prediction from a normally distribution with mean $\mathbf{x_i^*}\bm{\beta^*_k} + \mathbf{z_i} \bm{\eta^*_k} $ and variance $\sigma^{2*}_k$. If a subject is part of a new cluster that currently has no members, we generate $\bm{\theta}_i$ using the base distribution $P_{0 \theta}$.

The probability that a subject is in the $k^{th}$ existing $\theta$-cluster is proportional to:

\[
\frac{ n_k^{\theta} } {\alpha_{\theta} + n }  \times \left[ \frac{ \alpha_{\psi} } {\alpha_{\psi} + n_k^{\theta} }  f_{x,0}(\mathbf{x}_i) + \sum_j \left( \frac{n_{j|k}^{\psi}}{\alpha_{\psi} + n_k^{\theta}} \prod_{l=1}^p f_{x, l}(x_{i, l}; \bm{\psi}^*_{j|k}) \right)\right],
\]
where the summation iterates through all nested $\psi$-clusters for the $k^{th}$ $\theta$-cluster.

The probability that a subject is in a new $\theta$-cluster is proportional to:

\[
\frac{ \alpha_{\theta} } {\alpha_{\theta} + n }  \times  f_{x,0}(\mathbf{x}_i),
\]
where $f_{x, 0}(\mathbf{x}_i) = \prod_{l = 1}^p \int_\psi f_{x, l}(x_{i,l})dP_{0\psi|\theta}$, the density integrated over the base measure evaluated at the observed data \cite{wade2014improving}. This computation for binary and continuous covariates using our distributional and prior assumptions is shown in the appendix. Since we used conjugate priors, this integration can be done analytically. When non-conjugate priors are used, Monte Carlo integration is an option.

\section{Simulations}

We used simulation to assess the predictive performance of our longitudinal model with splines and an EDP prior. For each simulated subject, we predicted the outcome at a specific time and compared it to the true value. Let $y_{i,t}$ be the $i^{th}$ subject's true value at time $t$ and let $\hat{y}_{i,t}$ be the prediction of $y_{i,t}$ from a given model. We computed the mean absolute prediction error $L_1$ and the mean squared prediction error $L_2$ over all simulated subjects.

\begin{align*}
\ell_1 &= \frac{1}{n} \sum_{i = 1}^n \left|\hat{y}_{i,t} - y_{i,t}\right| \\
\ell_2 &= \frac{1}{n} \sum_{i = 1}^n \left(\hat{y}_{i,t} - y_{i,t}\right)^2
\end{align*}

\begin{figure}
	\centering
	\includegraphics[width=1.0\textwidth]{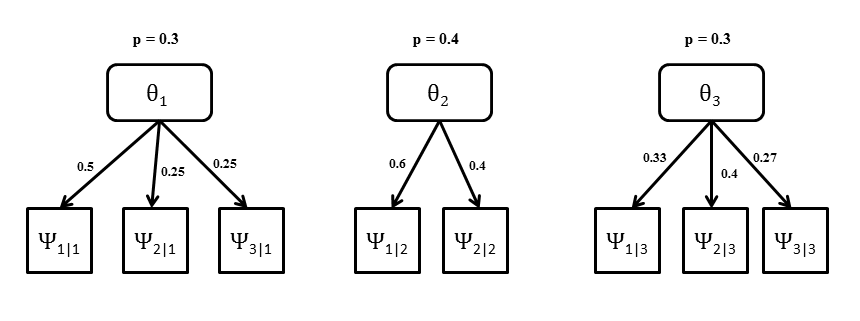}
	\caption{Structure of clustering for simulations with probabilities of being in each cluster. Probabilities for being in a $\psi$-cluster are conditional on being in the appropriate $\theta$-cluster. $\theta$ refers to the regression parameters and $\psi$ refers to the covariate parameters.}
	\label{fig:clust}
\end{figure}

We simulated sample sizes of $n = 1000$ and $n = 5000$. Each individual was randomly assigned a minimum of 1 and a maximum of 5 repeated measurements corresponding to time points within the interval $t \in [0, 1]$ generated randomly from an independent uniform distribution. As before, let $\theta$ denote the regression parameters and $\psi$ denote the covariate parameters. The true cluster structure had three $\theta$-clusters. Within each $\theta$-cluster, there were 3, 2, and 3 nested $\psi$-clusters. Thus, the total number of unique clusters was 8 while the total number of unique $\theta$-clusters was 3. The structure of the clustering along with probabilities of being in each cluster are given in Figure~\ref{fig:clust}. Each subject was assigned 20 simulated covariates from distributions whose parameters differed between $\psi$-clusters. Full data-generating details are available in the appendix and code is available upon request (code for the EDP and DP models are at https://www.github.com/zeldow/EDPlong and https://www.github.com/zeldow/DPlong).


Predictions were made for each subject at $t = 0.75$ and the true value $y_{i, t}$ was calculated based on the mean for the $\theta$-cluster to which the individual belongs (mean function shown in appendix) and the random intercept. We generated 100 datasets and take the mean of $\ell_1$ and $\ell_2$ over all simulations, and then calculate

\begin{align*}
\bar{\ell}_1 & = \frac{1}{100} \sum_{j=1}^{100} \ell_{1j} \\ 
\bar{\ell}_2 & = \frac{1}{100} \sum_{j=1}^{100} \ell_{2j},
\end{align*}
where $\ell_{1j}$ and $\ell_{2j}$ are $\ell_1$ and $\ell_2$ calculated on the $j^{th}$ simulated dataset.

To assess the performance of our mixed model with an EDP prior, we compared it to two competitor models: a Bayesian mixed model with a DP prior and a linear mixed model (implemented by the lme4 package \cite{bates2014lme4} in R \cite{rproj}). For each sample size, we varied the regression variance $\sigma^2$ and the random intercept variance $\sigma^2_u$ resulting in four simulation scenarios: (1) low variability in $\sigma^2$ and low variability in $\sigma^2_u$; (2) low variability in $\sigma^2$ and high variability in $\sigma^2_u$; (3) high variability in $\sigma^2$ and low variability in $\sigma^2_u$; and (4) high variability in $\sigma^2$ and high variability in $\sigma^2_u$. All models were fit using thin plate splines for the time effect. In the appendix, we show results using cubic B-splines as well with 2 knots at $\frac{1}{3}$ and $\frac{2}{3}$.

\begin{table}
\centering
\begin{tabular} {l c c c c c c}
\hline
& \multicolumn{2}{ c } {EDP} & \multicolumn{2}{ c } {DP} &  \multicolumn{2}{ c } {ME} \\
& $\bar{\ell}_1$ & $\bar{\ell}_2$ & $\bar{\ell}_1$ & $\bar{\ell}_2$ & $\bar{\ell}_1$ & $\bar{\ell}_2$ \\
\hline
$\sigma^2 = 1$; $\sigma^2_u = 0.15$ & 0.66 & 0.87 & 0.89 & 1.43 & 1.11  & 1.85 \\
$\sigma^2 = 1$; $\sigma^2_u = 0.5$ & 0.82 & 1.19 & 1.07 & 1.93 & 1.11 & 1.87 \\
$\sigma^2 = 4$; $\sigma^2_u = 0.15$ & 0.89 & 1.46 & 1.08 & 2.00 & 1.23 & 2.32 \\
$\sigma^2 = 4$; $\sigma^2_u = 0.5$ & 1.05 & 1.90 & 1.17 & 2.28 & 1.24 & 2.37 \\
\hline
\end{tabular}
\caption{Simulation results for $n = 1000$ showing mean $l_1$ and $l_2$ errors over 100 datasets for predictions at $t = 0.75$. $\sigma^2$ indicates the simulated regression variance and $\sigma^2_u$ indicates the simulated random intercept variance. EDP indicates the longitudinal model with an enriched Dirichlet process prior. DP indicates the longitudinal model with a Dirichlet process prior. ME indicates a mixed effects model fit using the lmer package in R \cite{bates2005fitting}. Fit with penalized thin plate splines with 20 knots.}
\label{tab:sim1}
\end{table}

\begin{table}
\centering
\begin{tabular} {l c c c c c c}
\hline
& \multicolumn{2}{ c } {EDP} & \multicolumn{2}{ c } {DP} &  \multicolumn{2}{ c } {ME} \\
& $\bar{\ell}_1$ & $\bar{\ell}_2$ & $\bar{\ell}_1$ & $\bar{\ell}_2$ & $\bar{\ell}_1$ & $\bar{\ell}_2$ \\
\hline
$\sigma^2 = 1$; $\sigma^2_u = 0.15$ & 0.60 & 0.73 & 0.91 & 1.49 & 1.10 & 1.82 \\
$\sigma^2 = 1$; $\sigma^2_u = 0.5$ & 0.77 & 1.06 & 1.10 & 2.03 & 1.11 & 1.86 \\
$\sigma^2 = 4$; $\sigma^2_u = 0.15$ & 0.71 & 0.99 & 1.04 & 1.90 & 1.21 & 2.27 \\
$\sigma^2 = 4$; $\sigma^2_u = 0.5$ & 0.85 & 1.27 &1.15 & 2.25 & 1.23 & 2.34  \\
\hline
\end{tabular}
\caption{Simulation results for $n = 5000$ showing mean $l_1$ and $l_2$ errors over 100 datasets for predictions at $t = 0.75$. $\sigma^2$ indicates the simulated regression variance and $\sigma^2_u$ indicates the simulated random intercept variance. EDP indicates the longitudinal model with an enriched Dirichlet process prior. DP indicates the longitudinal model with a Dirichlet process prior. ME indicates a mixed effects model fit using the lmer package in R \cite{bates2005fitting}. Fit with penalized thin plate splines with 20 knots.} 
\label{tab:sim2}
\end{table}

The results of the simulation study for $n = 1000$ are shown in Table~\ref{tab:sim1}. For all scenarios the EDP model outperformed the DP model and the mixed model in terms of mean $L_1$ and $L_2$ prediction error. The mean $L_1$ error for the EDP model ranged from 0.66 to 1.05. For the DP model, it ranged from 0.89 to 1.17, and for the standard mixed model it ranged from 1.11 to 1.24. The mean $L_2$ errors range from 0.87 to 1.90, 1.43 to 2.28, and 1.85 to 2.37 among the models for the four simulation scenarios, respectively. Given that the data were generated in clusters, it is unsurprising that the two methods implementing clustering provide better predictions than the standard mixed effects model. However, we see that the EDP model yielded more precise prediction than the DP model based on $L_1$ and $L_2$ prediction error.

The results in Table~\ref{tab:sim2} display the results for the simulations with $n = 5000$. Again, the EDP model outperforms the DP model which outperforms the mixed model. For the most part, there are no large differences in the relative performance of methods at the two sample sizes. Using cubic B-splines (see Appendix) in lieu of thin plate splines also did not have considerable effect on prediction error. Overall the results for thin-plate splines were slightly improved over those of B-splines, but more research needs to be done and more scenarios examined. 

\begin{table}[!ht]
\centering
\begin{tabular} {l c c c c c c}
\hline
& \multicolumn{2}{ c } {EDP} & \multicolumn{2}{ c } {DP} &  \multicolumn{2}{ c } {ME} \\
& $\bar{\ell}_1$ & $\bar{\ell}_2$ & $\bar{\ell}_1$ & $\bar{\ell}_2$ & $\bar{\ell}_1$ & $\bar{\ell}_2$ \\
\hline
$\sigma^2 = 1$; $\sigma^2_u = 0.15$ & 0.31 & 0.16 & 0.31 & 0.16 & 0.28  & 0.12 \\
$\sigma^2 = 1$; $\sigma^2_u = 0.5$ & 0.56 & 0.50 & 0.56 & 0.50 & 0.38 & 0.23 \\
$\sigma^2 = 4$; $\sigma^2_u = 0.15$ & 0.34 & 0.18 & 0.34 & 0.18 & 0.35 & 0.20 \\
$\sigma^2 = 4$; $\sigma^2_u = 0.5$ & 0.58 & 0.52 & 0.58 & 0.52 & 0.52 & 0.42 \\
\hline
\end{tabular}
\caption{Simulation results for $n = 1000$ showing mean $l_1$ and $l_2$ errors over 100 datasets for predictions at $t = 0.75$. $\sigma^2$ indicates the simulated regression variance and $\sigma^2_u$ indicates the simulated random intercept variance. EDP indicates the longitudinal model with an enriched Dirichlet process prior. DP indicates the longitudinal model with a Dirichlet process prior. ME indicates a mixed effects model fit using the lmer package in R \cite{bates2005fitting}. Fit with penalized thin plate splines with 20 knots.}
\label{tab:sim3}
\end{table}

Lastly, we performed simulations where all subjects were part of the same cluster so that the linear mixed model was correctly specified and was expected to work best. The results for $n=1000$ are included in Table~\ref{tab:sim3}. Over 100 simulated datasets, the correctly specified standard mixed model outperformed both the EDP and DP models in almost all scenarios. Results from EDP and DP models showed no difference up to two decimal places. This interesting finding was due to the fact that the EDP model did not split $\theta$-clusters in subclusters, rendering the difference between the DP and EDP models irrelevant. Overall, we found that the EDP and DP models concentrated around one large $\theta$-cluster with scattered observations in other $\theta$-clusters. With $n = 1000$, the $L_1$ error for the mixed model ranged from 0.28 to 0.52, while for the DP and EDP models, it ranged from 0.31 to 0.58. The largest difference in favor of the standard mixed model occurred with low regression variance $\sigma^2 = 1$ and high random intercept variance $\sigma_u^2 = 0.5$ ($L_1$ error: 0.38 versus 0.56; $L_2$ error: 0.23 versus 0.50). Other scenarios showed either no difference or only a modest improvement for the standard mixed model. One possible explanation for this discrepancy is that there is an identifiability problem in the models with DP or EDP priors in which the algorithm has difficulty determining if $\sigma^2_u$ is smaller and there are many clusters or if $\sigma^2_u$ is large and there are few clusters. Thus, in this scenario the EDP and DP models split the sample into more clusters than was necessary and prediction suffered accordingly. On the other hand, the reverse scenario with high regression variance and low random intercept variance showed no difference between the three models, indicating that the nonparametric prior performed fine in more likely situations.

\section{Data Analysis}

Sentinel is an initiative of the US  Food and Drug Administration with 17 data partners \cite{sentinel}. Under Sentinel, a distributed database has been established that collects EHR and administrative health plan data to assess safety in approved medical products, particularly drugs and vaccines. As part of a workgroup effort to understand and use laboratory results data in the Sentinel Distributed Database (SDD) \cite{analytic}, \citet{flory2017missing} used the SDD to calculate incidence rates of diabetes among new initiators of second generation antipsychotics (SGAs), which are known to increase the risk of Type II diabetes mellitis (T2DM) \cite{newcomer2005second, de2012metabolic}. T2DM is often diagnosed based on elevated levels of hemoglobin A1c (HbA1c), serum glucose, or capillary glucose \cite{american2014diagnosis}. In \citet{flory2017missing} incidence rates for T2DM were computed from two outcomes: (O1) diagnosis codes and dispensement of antidiabetic medication and (O2) diagnosis codes, dispensement of antidiabetic medication as well as an elevated diabetes labs. Lab values were considered elevated if fasting glucose $\geq 126$ mg/dl, random glucose $\geq 200$ mg/dl, or HbA1c $\geq6.5$v\%. Including diabetes labs increased the number of T2DM cases, but missingness was differential among the sites analyzed, affecting some sites more than others. In this paper, we extend some of the results of \citet{flory2017missing} using predictions from our longitudinal EDP model.

We restricted our analysis to site one of \citet{flory2017missing}, which corresponds to a small integrated delivery system. As in that publication, our cohort was restricted to participants at least 21 years of age who had at least 183 days of health plan enrollment prior to initiating a SGA (aripiprazole, olanzapine, quetiapine, and risperidone). We included those who had first dispensement of a SGA between 1 January 2008 and 31 October 2012. Any individuals with evidence of diabetes prior to initiation of the SGA, including diagnosis of diabetes, receipt of an antidiabetic medication, or an elevated diabetes lab, were excluded. Follow-up began at first dispensement of a SGA and continued until discontinuation of insurance, death, occurrence of the outcome, or end of 365 days, whichever came first. The outcome was incident diabetes within 365 days of study, equal to that of outcome O1 above. We also define a new outcome O3, which consists of O1 and predicted elevated lab values.

The motivation for using our EDP longitudinal model for this problem is as follows. Our interest lies in calculating the incidence of diabetes within one year of initiating a SGA, supplementing the outcome with information from recorded lab values. The previous analysis was limited by restricting to lab values within one year of follow-up. However, lab values after one year can be informative as well, particularly those drawn soon after study end. Over 30\% of the subjects from site one did not have any lab values recorded between 1 and 365 days of SGA initiation. Subjects with lab values recorded had differential amounts of data recorded within that study window, ranging from 1 to 4 records for HbA1c, 1 to 5 of fasting glucose, and 1 to 115 of random glucose. Lastly, the approach in \citet{flory2017missing} treats any instance of a lab value exceeding the threshold as part of the outcome even if only one measurement among many exceeded the threshold. Because of this, uncertainty stemming from measurement error was inadequately accounted for. Our model incorporates such uncertainty through the regression variance component $\sigma^2$ as well as the fact that cluster membership $s_y$ changes throughout the algorithm.

We fit EDP longitudinal models for each of three lab values (HbA1c, fasting glucose, and random glucose) separately. Models were fit with the entire history of the subject's lab values until initiation of an anti-diabetic medication. Our dataset had a total of $n = 3,764$ study participants. Among these, 680 subjects contributed 1,003 observations for HbA1c. For fasting glucose, 2,032 subjects contributed 4,110 observations. For random glucose, 3,013 subjects contributed 21,614 observations. We used 200,000 iterations with 40,000 burn in period. Throughout the 160,000 post burn in iterations, predictions were drawn at 800 evenly spaced iterations. Each subject had predictions made at day 365, unless their study censoring time was prior to that, at which point we made predictions at that censoring time. All predicted values were appended to the original dataset resulting in 800 imputed datasets. Each imputed dataset consists of the original data, including diabetes diagnoses and dispensement of antidiabetics, along with three predicted values for HbA1c, random glucose, and fasting glucose. The outcome O3 was calculated for each imputed dataset. From this, we then calculate the incidence of diabetes and use multiple imputation methods to combine estimates across imputations \cite{rubin2004multiple}. Overall, the HbA1c model took 4.7 hours of runtime, the fasting glucose model 22.4 hours, and the random glucose model 63.2 hours.

In total, 89 participants were diagnosed with diabetes through diagnosis codes or dispensement of anti-diabetic medication. The total number of outcomes O3 ranged from 146 to 394 outcomes with a median of 200 throughout the 800 imputations. This resulted in an incidence of 0.059 events per person-year (95\% confidence interval: 0.043--0.080). This result is similar to the incidence found in \citet{flory2017missing} for site one among those with recorded lab values, except the confidence interval is wider, reflecting greater uncertainty in classification using lab values.

\subsection{Clustering}

We also examined clustering resulting from our model. There is a multitude of reasons one may be interested in clustering in the present example. First, it can show heterogeneity (or lack thereof) of outcome features among groups of individuals. The cluster itself may be able to predict outcomes. For example, if we know that a certain individual is in a cluster with rising HbA1c values over time, we know that their likelihood of a diabetes diagnosis is increased compared to a group with flat trajectories over time. Further, once we have identified the clustering structure, we can examine the distributions of covariates within cluster and determine covariates that may be affecting the differences among groups.

Recall that we refer to clusters based on regression parameters as $\theta$-clusters and the nested clusters based on covariates as $\psi$-clusters. For illustrative purposes, we focus strictly on functional clustering using $\theta$-clusters. Other applications may have interest in summarizing $\psi$-clusters as well. Given that within the MCMC algorithm, not only cluster membership but the number of clusters can change, we condense the results into a single point estimate for the posterior cluster structure. For the HbA1c and fasting glucose models, the posterior number of clusters concentrated around two. For random glucose, the posterior number of clusters concentrated around three. All models were initialized to have two $\theta$-clusters. When we initialized the number of $\theta$-clusters to 10, results eventually converged to similar answers for each of the outcomes. However, computation time was considerably longer when initialized with a large number of $\theta$-clusters.

\begin{figure}
    \centering
    \includegraphics[width=1\textwidth]{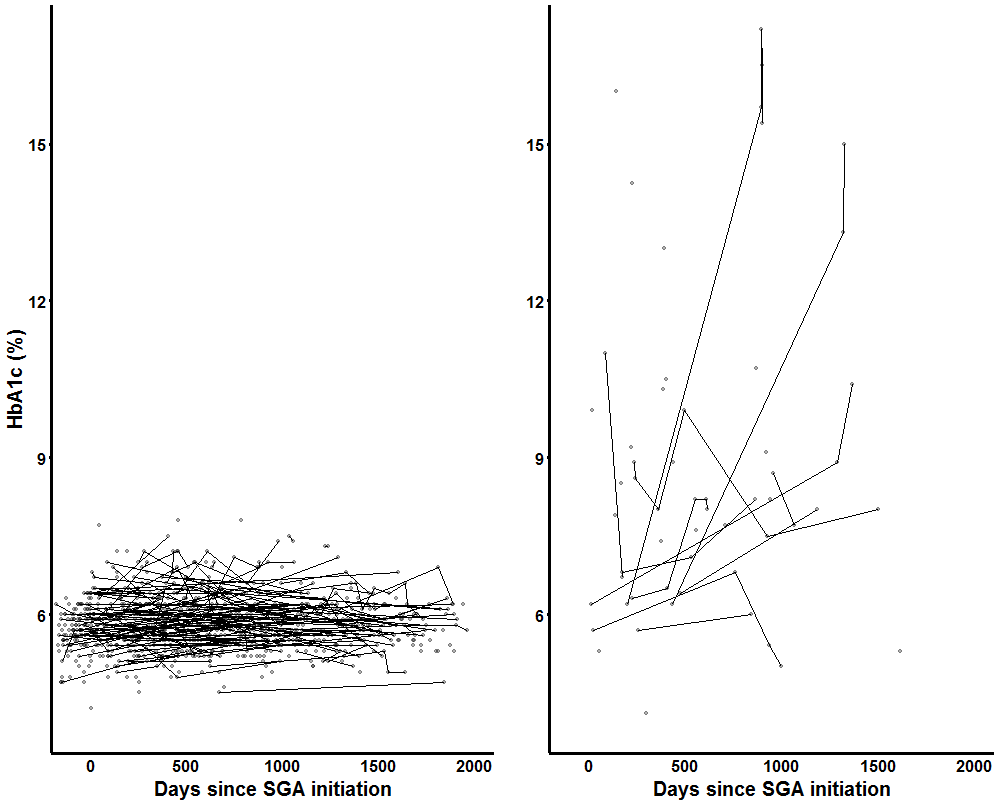}
    \caption{Clustering results for HbA1c model. The model settled on two $\theta$-clusters which are shown in the figure. The larger cluster has 650 subjects and the smaller cluster has 30 subjects.}
  \label{fig:a1c}
\end{figure}

\begin{figure}
    \centering
    \includegraphics[width=1\textwidth]{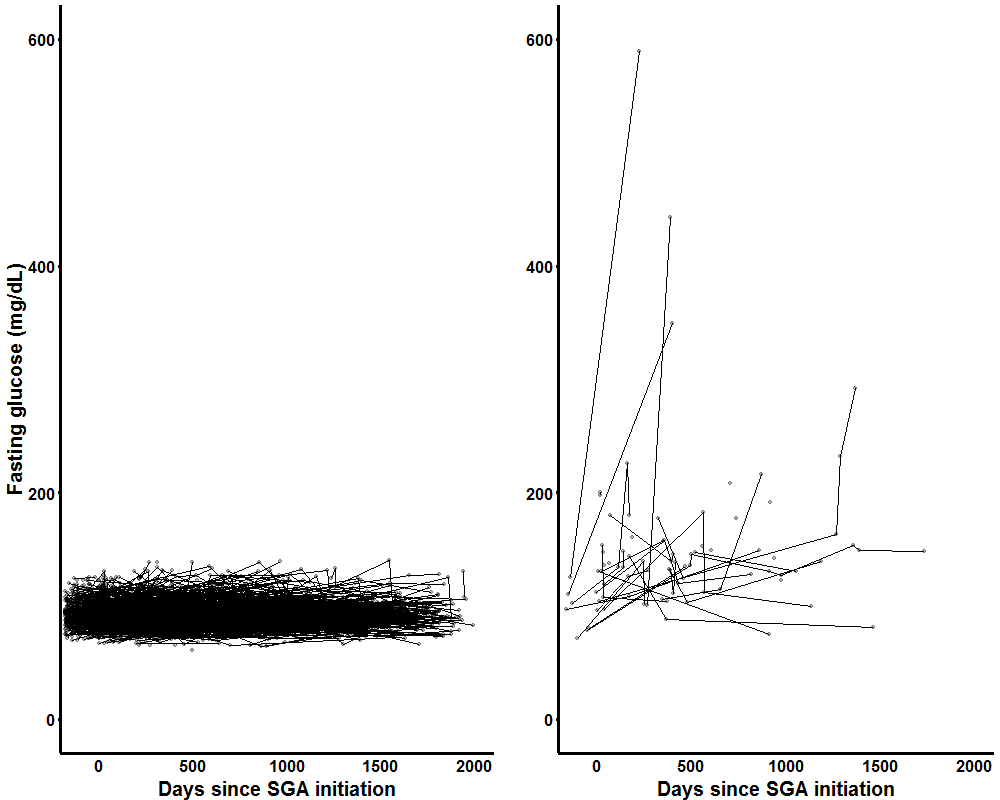}
    \caption{Clustering results for fasting glucose model. The model settled on two $\theta$-clusters. The larger cluster has 1997 subjects and the smaller cluster has 35 subjects.}
  \label{fig:fast}
\end{figure}

\begin{figure}
    \centering
    \includegraphics[width=1\textwidth]{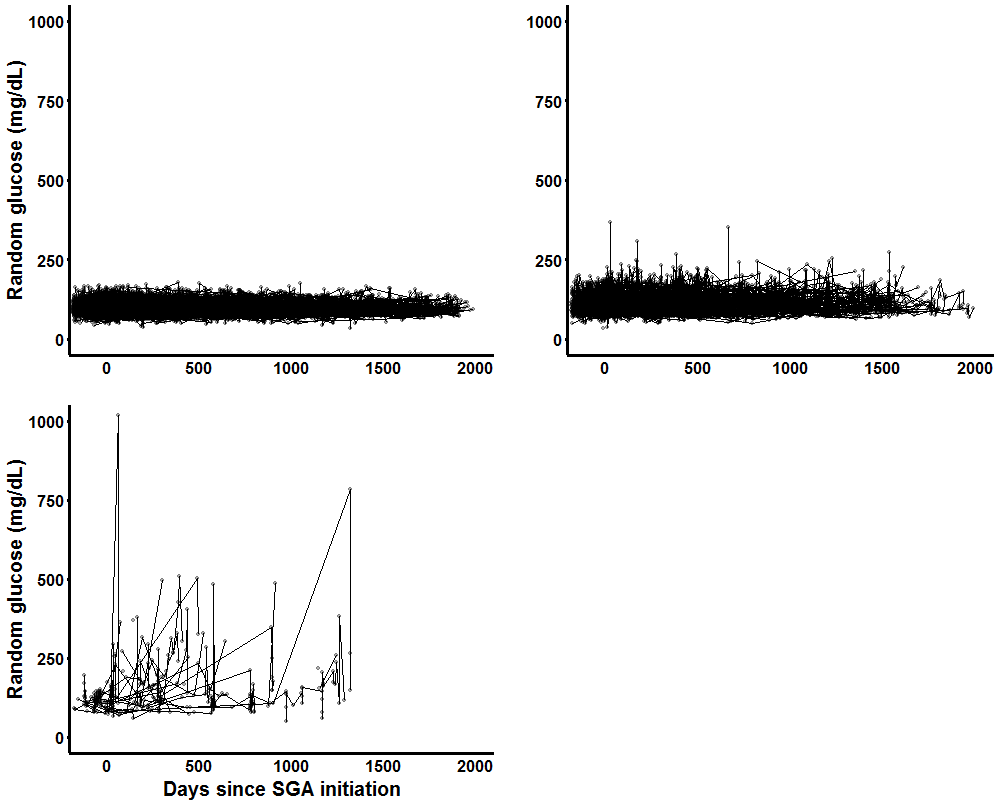}
    \caption{Clustering results for random glucose model. The model settled on three $\theta$-clusters with 2563, 419, and 31 subjects.}
  \label{fig:rand}
\end{figure}

For both the model with HbA1c as the outcome and the model with fasting glucose as the outcome, our algorithm settled on two distinct clusters as seen in Figure~\ref{fig:a1c} and Figure~\ref{fig:fast}, respectively. For HbA1c, the first cluster contained 650 observations consisting of trajectories that mostly stay within the values 5\% and 8\%. The remaining 30 observations in the second cluster consisted of highly variable trajectories that had spikes in their values. In the fasting glucose model, the first cluster had 1997 members and consisted of tight trajectories below the threshold of 126 mg/dL, while the second cluster housed the remaining 35 subjects mostly of subjects whose trajectory at some point contains a spike or is somehow indicative of higher variability. 

The model with random glucose as the outcome settled on three clusters which can be seen in Figure~\ref{fig:rand}. The largest cluster had 2563 subjects who had relatively flat trajectories with small within-subject variability. The second largest cluster contained 419 subjects who had trajectories with slightly more variability than the first cluster. The third cluster contains 31 subjects with large spikes and characterized by larger variability than the other clusters.

\section{Discussion}

In this paper, we presented a joint model for a continuous longitudinal outcome and the baseline covariates. The model is partitioned into the product of a linear mixed model for the outcome given the covariates and the marginal distributions for the covariates. The use of the EDP prior in a longitudinal model is an extension of the model developed by \citet{wade2011enriched}, which itself is an extension of the DP prior. Through the nested clustering of the EDP prior, where subjects are clustered separately for their regression trajectories and similarity in the covariate space, our model allows for improved prediction over the same model with the usual DP prior. This improvement was demonstrated in simulation scenarios in which the EDP longitudinal model outperformed both a standard mixed model and a longitudinal model with a DP prior when the data generating distribution contained a nested clustering structure. When the simulation scenario was simplified so that there was no underlying cluster structure and the linear mixed model was correctly specified, using the nonparametric EDP prior did not excessively diminish predictive performance. Our model also serves as a functional clustering algorithm, the first to use an EDP prior. In our model setup, the EDP prior is particularly useful because it allows the functional to cluster solely on functional features rather than non-functional components (i.e., closeness in the covariate space).


One limitation of the present model is that it can only incorporate baseline covariates. In many longitudinal settings, covariates may be updated throughout the study. One possibility to incorporate this into our model would be to use the dynamic DP, which allows for distributions to evolve in discrete time \cite{rodriguez2008bayesian}. From the current state of the literature, DPs which evolve throughout time are less thoroughly developed and more difficult to implement. The extension of our model to handle time-varying covariates is a topic for future research.

Throughout the paper we made several modeling choices that could be changed or generalized. For example, the value for $\alpha_{\psi}$ could depend on $\theta$ so that the mass parameter is written as $\alpha_{\psi}(\theta)$. This would allow the number of subclusters to differ depending on the value of $\theta$. Further, we made the assumption that the values of $\psi$ and $\theta$ were independent through the fact that $P_0 = P_{0 \theta} \times P_{0 \psi |  \theta}$. This assumption simplifies calculations but can be relaxed if needed. These two changes were discussed in \citet{wade2011enriched}. Lastly, we focused on continuous outcomes, but our methods can be extended to more general settings such as binary or count outcomes or different link functions with various additional computational challenges (non-conjugacy for one). Bayesian computations for generalized linear mixed models are provided in \citet{zhao2006general} and references therein.

We demonstrated our model with data from the Sentinel Distributed Database, where we used predicted lab measurements to augment incidence rates of diabetes among subjects initiating certain anti-psychotics. Our incidence rates were similar to those found in a previous paper on the same study population \cite{flory2017missing}, but our estimates gave wider confidence intervals, reflecting greater uncertainty about the incorporation of labs as part of the diabetes diagnosis. Our model is well-suited for other applications as well, such as with data arising from studies using wearable devices or studies of symptoms of chronic conditions with interest in detecting patterns among patients.

\section*{Acknowledgements}

The authors gratefully acknowledge Christine Lu, Joshua Gagne, Azadeh Shoaibi, Lisa Herrinton, Mano Selvan, Xingmei Wang, and Elisabetta Patorno for their work developing the research questions, designing the original study, and creating the analytic database. This project was supported in part by Task Order HHSF22301012T-0008 under Master Agreement HHSF2232009100061 from the US Food and Drug Administration (FDA). The FDA approved the study protocol including statistical analysis plan, and reviewed and approved this manuscript. Coauthors from the FDA participated in the results interpretation and in the preparation and decision to submit the manuscript for publication. The FDA had no role in data collection, management, or analysis.

\bibliographystyle{unsrtnat}
\bibliography{paper2_arxiv_06JUN2018}

\begin{thebibliography}{41}
\providecommand{\natexlab}[1]{#1}
\providecommand{\url}[1]{\texttt{#1}}
\expandafter\ifx\csname urlstyle\endcsname\relax
  \providecommand{\doi}[1]{doi: #1}\else
  \providecommand{\doi}{doi: \begingroup \urlstyle{rm}\Url}\fi

\bibitem[of~Sciences{,}~Engineering{, and
  }Medicine(2017)]{national2017refining}
National~Academies of~Sciences{,}~Engineering{, and }Medicine.
\newblock \emph{Refining the Concept of Scientific Inference When Working with
  Big Data: Proceedings of a Workshop}.
\newblock National Academies Press, 2017.

\bibitem[Association(2014)]{american2014diagnosis}
American~Diabetes Association.
\newblock Diagnosis and classification of diabetes mellitus.
\newblock \emph{Diabetes care}, 37\penalty0 (Supplement 1):\penalty0 S81--S90,
  2014.

\bibitem[Wade et~al.(2011)Wade, Mongelluzzo, and Petrone]{wade2011enriched}
Sara Wade, Silvia Mongelluzzo, and Sonia Petrone.
\newblock An enriched conjugate prior for bayesian nonparametric inference.
\newblock \emph{Bayesian Analysis}, 6\penalty0 (3):\penalty0 359--385, 2011.

\bibitem[Ferguson(1973)]{ferguson1973bayesian}
Thomas~S Ferguson.
\newblock A bayesian analysis of some nonparametric problems.
\newblock \emph{The annals of statistics}, pages 209--230, 1973.

\bibitem[Ferguson(1983)]{ferguson1983bayesian}
Thomas~S Ferguson.
\newblock Bayesian density estimation by mixtures of normal distributions.
\newblock \emph{Recent advances in statistics}, 24\penalty0 (1983):\penalty0
  287--302, 1983.

\bibitem[Escobar and West(1995)]{escobar1995bayesian}
Michael~D Escobar and Mike West.
\newblock Bayesian density estimation and inference using mixtures.
\newblock \emph{Journal of the american statistical association}, 90\penalty0
  (430):\penalty0 577--588, 1995.

\bibitem[Teh et~al.(2004)Teh, Jordan, Beal, and Blei]{teh2004sharing}
Yee~Whye Teh, Michael~I Jordan, Matthew~J Beal, and David~M Blei.
\newblock Sharing clusters among related groups: Hierarchical dirichlet
  processes.
\newblock In \emph{NIPS}, pages 1385--1392, 2004.

\bibitem[Hanson and Johnson(2004)]{hanson2004bayesian}
Timothy Hanson and Wesley~O Johnson.
\newblock A bayesian semiparametric aft model for interval-censored data.
\newblock \emph{Journal of Computational and Graphical Statistics}, 13\penalty0
  (2):\penalty0 341--361, 2004.

\bibitem[Hannah et~al.(2011)Hannah, Blei, and Powell]{hannah2011dirichlet}
Lauren~A Hannah, David~M Blei, and Warren~B Powell.
\newblock Dirichlet process mixtures of generalized linear models.
\newblock \emph{Journal of Machine Learning Research}, 12\penalty0
  (Jun):\penalty0 1923--1953, 2011.

\bibitem[Cruz-Mes{\'\i}a et~al.(2007)Cruz-Mes{\'\i}a, Quintana, and
  M{\"u}ller]{cruz2007semiparametric}
Rolando De~la Cruz-Mes{\'\i}a, Fernando~A Quintana, and Peter M{\"u}ller.
\newblock Semiparametric bayesian classification with longitudinal markers.
\newblock \emph{Journal of the Royal Statistical Society: Series C (Applied
  Statistics)}, 56\penalty0 (2):\penalty0 119--137, 2007.

\bibitem[Roy et~al.(2017)Roy, Lum, Daniels, Zeldow, Dworkin, and
  Re~III]{roy2017bayesian}
Jason Roy, Kirsten~J Lum, Michael~J Daniels, Bret Zeldow, Jordan Dworkin, and
  Vincent~Lo Re~III.
\newblock Bayesian nonparametric generative models for causal inference with
  missing at random covariates.
\newblock \emph{arXiv preprint arXiv:1702.08496}, 2017.

\bibitem[Shahbaba and Neal(2009)]{shahbaba2009nonlinear}
Babak Shahbaba and Radford Neal.
\newblock Nonlinear models using dirichlet process mixtures.
\newblock \emph{Journal of Machine Learning Research}, 10\penalty0
  (Aug):\penalty0 1829--1850, 2009.

\bibitem[McCullagh(1984)]{mccullagh1984generalized}
Peter McCullagh.
\newblock Generalized linear models.
\newblock \emph{European Journal of Operational Research}, 16\penalty0
  (3):\penalty0 285--292, 1984.

\bibitem[Gelman et~al.(2014)Gelman, Carlin, Stern, and
  Rubin]{gelman2014bayesian}
Andrew Gelman, John~B Carlin, Hal~S Stern, and Donald~B Rubin.
\newblock \emph{Bayesian data analysis}, volume~2.
\newblock Chapman \& Hall/CRC Boca Raton, FL, USA, 2014.

\bibitem[M{\"u}ller et~al.(2015)M{\"u}ller, Quintana, Jara, and
  Hanson]{muller2015bayesian}
Peter M{\"u}ller, Fernando~Andr{\'e}s Quintana, Alejandro Jara, and Tim Hanson.
\newblock \emph{Bayesian nonparametric data analysis}.
\newblock Springer, 2015.

\bibitem[Wade et~al.(2014)Wade, Dunson, Petrone, and Trippa]{wade2014improving}
Sara Wade, David~B Dunson, Sonia Petrone, and Lorenzo Trippa.
\newblock Improving prediction from dirichlet process mixtures via enrichment.
\newblock \emph{Journal of Machine Learning Research}, 15\penalty0
  (1):\penalty0 1041--1071, 2014.

\bibitem[M{\"u}ller and Rosner(1997)]{muller1997bayesian}
Peter M{\"u}ller and Gary~L Rosner.
\newblock A bayesian population model with hierarchical mixture priors applied
  to blood count data.
\newblock \emph{Journal of the American Statistical Association}, 92\penalty0
  (440):\penalty0 1279--1292, 1997.

\bibitem[Li et~al.(2010)Li, Lin, and M{\"u}ller]{li2010bayesian}
Yisheng Li, Xihong Lin, and Peter M{\"u}ller.
\newblock Bayesian inference in semiparametric mixed models for longitudinal
  data.
\newblock \emph{Biometrics}, 66\penalty0 (1):\penalty0 70--78, 2010.

\bibitem[Das et~al.(2013)Das, Li, Sengupta, and Wu]{das2013bayesian}
Kiranmoy Das, Runze Li, Subhajit Sengupta, and Rongling Wu.
\newblock A bayesian semiparametric model for bivariate sparse longitudinal
  data.
\newblock \emph{Statistics in medicine}, 32\penalty0 (22):\penalty0 3899--3910,
  2013.

\bibitem[Quintana et~al.(2016)Quintana, Johnson, Waetjen, and
  B.~Gold]{quintana2016bayesian}
Fernando~A Quintana, Wesley~O Johnson, L~Elaine Waetjen, and Ellen B.~Gold.
\newblock Bayesian nonparametric longitudinal data analysis.
\newblock \emph{Journal of the American Statistical Association}, 111\penalty0
  (515):\penalty0 1168--1181, 2016.

\bibitem[Bigelow and Dunson(2009)]{bigelow2009bayesian}
Jamie~L Bigelow and David~B Dunson.
\newblock Bayesian semiparametric joint models for functional predictors.
\newblock \emph{Journal of the American Statistical Association}, 104\penalty0
  (485):\penalty0 26--36, 2009.

\bibitem[Scarpa and Dunson(2014)]{scarpa2014enriched}
Bruno Scarpa and David~B Dunson.
\newblock Enriched stick-breaking processes for functional data.
\newblock \emph{Journal of the American Statistical Association}, 109\penalty0
  (506):\penalty0 647--660, 2014.

\bibitem[Jacques and Preda(2014)]{jacques2014functional}
Julien Jacques and Cristian Preda.
\newblock Functional data clustering: a survey.
\newblock \emph{Advances in Data Analysis and Classification}, 8\penalty0
  (3):\penalty0 231--255, 2014.

\bibitem[Newcomer(2005)]{newcomer2005second}
John~W Newcomer.
\newblock Second-generation (atypical) antipsychotics and metabolic effects.
\newblock \emph{CNS drugs}, 19\penalty0 (1):\penalty0 1--93, 2005.

\bibitem[De~Hert et~al.(2012)De~Hert, Detraux, Van~Winkel, Yu, and
  Correll]{de2012metabolic}
Marc De~Hert, Johan Detraux, Ruud Van~Winkel, Weiping Yu, and Christoph~U
  Correll.
\newblock Metabolic and cardiovascular adverse effects associated with
  antipsychotic drugs.
\newblock \emph{Nature Reviews Endocrinology}, 8\penalty0 (2):\penalty0
  114--126, 2012.

\bibitem[Flory et~al.(2017)Flory, Roy, Gagne, Haynes, Herrinton, Lu, Patorno,
  Shoaibi, and Raebel]{flory2017missing}
James~H Flory, Jason Roy, Joshua~J Gagne, Kevin Haynes, Lisa Herrinton,
  Christine Lu, Elisabetta Patorno, Azadeh Shoaibi, and Marsha~A Raebel.
\newblock Missing laboratory results data in electronic health databases:
  implications for monitoring diabetes risk.
\newblock \emph{Journal of Comparative Effectiveness Research}, 6\penalty0
  (1):\penalty0 25--32, 2017.

\bibitem[Rubin(2004)]{rubin2004multiple}
Donald~B Rubin.
\newblock \emph{Multiple imputation for nonresponse in surveys}, volume~81.
\newblock John Wiley \& Sons, 2004.

\bibitem[Hastie and Tibshirani(1990)]{hastie1990generalized}
Trevor~J Hastie and Robert~J Tibshirani.
\newblock \emph{Generalized additive models}, volume~43.
\newblock CRC press, 1990.

\bibitem[Crainiceanu et~al.(2005)Crainiceanu, Ruppert, and
  Wand]{crainiceanu2005bayesian}
Ciprian~M Crainiceanu, David Ruppert, and Matthew~P Wand.
\newblock Bayesian analysis for penalized spline regression using win bugs.
\newblock \emph{Journal of Statistical Software}, 14\penalty0 (14), 2005.

\bibitem[Ray and Mallick(2006)]{ray2006functional}
Shubhankar Ray and Bani Mallick.
\newblock Functional clustering by bayesian wavelet methods.
\newblock \emph{Journal of the Royal Statistical Society: Series B (Statistical
  Methodology)}, 68\penalty0 (2):\penalty0 305--332, 2006.

\bibitem[Medvedovic and Sivaganesan(2002)]{medvedovic2002bayesian}
Mario Medvedovic and Siva Sivaganesan.
\newblock Bayesian infinite mixture model based clustering of gene expression
  profiles.
\newblock \emph{Bioinformatics}, 18\penalty0 (9):\penalty0 1194--1206, 2002.

\bibitem[Murtagh and Legendre(2014)]{murtagh2014ward}
Fionn Murtagh and Pierre Legendre.
\newblock Ward’s hierarchical agglomerative clustering method: which
  algorithms implement ward’s criterion?
\newblock \emph{Journal of Classification}, 31\penalty0 (3):\penalty0 274--295,
  2014.

\bibitem[Neal(2000)]{neal2000markov}
Radford~M Neal.
\newblock Markov chain sampling methods for dirichlet process mixture models.
\newblock \emph{Journal of computational and graphical statistics}, 9\penalty0
  (2):\penalty0 249--265, 2000.

\bibitem[Zeger and Karim(1991)]{zeger1991generalized}
Scott~L Zeger and M~Rezaul Karim.
\newblock Generalized linear models with random effects; a gibbs sampling
  approach.
\newblock \emph{Journal of the American statistical association}, 86\penalty0
  (413):\penalty0 79--86, 1991.

\bibitem[Bates et~al.(2014)Bates, Maechler, Bolker, and Walker]{bates2014lme4}
Douglas Bates, Martin Maechler, Ben Bolker, and Steven Walker.
\newblock lme4: Linear mixed-effects models using eigen and s4.
\newblock \emph{R package version}, 1\penalty0 (7), 2014.

\bibitem[{R Core Team}(2017)]{rproj}
{R Core Team}.
\newblock \emph{R: A Language and Environment for Statistical Computing}.
\newblock R Foundation for Statistical Computing, Vienna, Austria, 2017.
\newblock URL \url{https://www.R-project.org/}.

\bibitem[Bates(2005)]{bates2005fitting}
Douglas Bates.
\newblock Fitting linear mixed models in r.
\newblock \emph{R news}, 5\penalty0 (1):\penalty0 27--30, 2005.

\bibitem[sen()]{sentinel}
Sentinel.
\newblock \url{https://www.sentinelinitiative.org/}.
\newblock Accessed: 2017-10-13.

\bibitem[Raebel et~al.()Raebel, Shetterly, Paolino, Lu, Gagne, Haynes, Flory,
  Patorno, Smith, Selvan, Herrinton, Harrell~Jr, Shoaibi, and Roy]{analytic}
Marsha~A Raebel, Susan Shetterly, Andrea~R Paolino, Christine~Y Lu, Joshua~J
  Gagne, Kevin Haynes, James Flory, Elisabetta Patorno, David~H Smith, Mano
  Selvan, Lisa~J Herrinton, Frank~E Harrell~Jr, Azadeh Shoaibi, and Jason Roy.
\newblock Analytic methods for using laboratory test results in active database
  surveillance.
\newblock
  \url{https://www.sentinelinitiative.org/sentinel/methods/analytic-methods-using-laboratory-test-results-active-database-surveillance}.
\newblock Accessed: 2017-10-13.

\bibitem[Rodriguez and Ter~Horst(2008)]{rodriguez2008bayesian}
Abel Rodriguez and Enrique Ter~Horst.
\newblock Bayesian dynamic density estimation.
\newblock \emph{Bayesian Analysis}, 3\penalty0 (2):\penalty0 339--365, 2008.

\bibitem[Zhao et~al.(2006)Zhao, Staudenmayer, Coull, and Wand]{zhao2006general}
Yihua Zhao, John Staudenmayer, Brent~A Coull, and Matthew~P Wand.
\newblock General design bayesian generalized linear mixed models.
\newblock \emph{Statistical Science}, pages 35--51, 2006.

\end{thebibliography}

\newpage

\begin{center}
{\large\bf SUPPLEMENTARY MATERIAL}
\end{center}

\section*{Appendix: Simulation data-generating distribution}

See Figure 2 for details on clustering structure. Each subject $i$ has a random number of time points observed drawn from a discrete uniform distribution on $[1, \dots, 5]$, call this $n_i$. We now randomly draw $n_i$ time points from a uniform distribution on $[0, 1]$ and order them as $t_{i1} < \dots < t_{in_i}$. For each subject we draw a random intercept from a $N(0, \sigma^2_u)$ distribution. The outcome is generated from independent normal distribution (given $u_i$) with variance $\sigma^2$ and the mean $\mu$ depending on which $\theta$-cluster the subject was randomly assigned to. For $\theta_1$, the outcome has mean $\mu = 2 + 7t - 2x_1 - 0.5 + 2 \cos(x_4) + u_i$. For $\theta_2$, $\mu = 7 - 20 (t - 0.4)^2 + 1.1 x_2 - 0.8 x_3 + 0.5 x_4^2 + u_i$. For $\theta_3$, $\mu = 6 - 8 (t - 0.75)^2 - 3 x_1 -  x_4 +  x_5 + u_i$. For all the above $t$ represents the randomly generated time points for each subject.

There were 20 covariates $x$ were generated as: 

\begin{multicols}{3}
\begin{align*}
\bm{\psi}_{1|1}:\\
x_1 &\sim Bern(0.5) \\
x_2 &\sim Bern(0.75) \\
x_3 &\sim Bern(0.2) \\
x_4 &\sim N(0,1) \\
x_5 &\sim N(\sqrt{2}, \sqrt{2}) \\
x_6 - x_{20} &\sim N(0, 1)
\end{align*}

\begin{align*}
\bm{\psi}_{2|1}:\\
x_1 &\sim Bern(0.3) \\
x_2 &\sim Bern(0.5) \\
x_3 &\sim Bern(0.5) \\
x_4 &\sim N(0.5, 0.5) \\
x_5 &\sim N(1, 2) \\
x_6 - x_{20}   &\sim N(-0.5, 1)
\end{align*}

\begin{align*}
\bm{\psi}_{3|1}:\\
x_1 &\sim Bern(0.5) \\
x_2 &\sim Bern(0.5) \\
x_3 &\sim Bern(0.8) \\
x_4 &\sim N(0.5,2) \\
x_5 &\sim N(0, 1) \\
x_6 - x_{20}  &\sim N(0.5, 1)
\end{align*}
\end{multicols}

\begin{multicols}{2}
\begin{align*}
\bm{\psi}_{1|2}:\\
x_1 &\sim Bern(0.75) \\
x_2 &\sim Bern(0.5) \\
x_3 &\sim Bern(0.35) \\
x_4 &\sim N(2,1) \\
x_5 &\sim N(0,1) \\
x_6 - x_{20} &\sim N(-0.5, 1)
\end{align*}

\begin{align*}
\bm{\psi}_{2|2}:\\
x_1 &\sim Bern(0.5) \\
x_2 &\sim Bern(0.5) \\
x_3 &\sim Bern(0.5) \\
x_4 &\sim N(1,2) \\
x_5 &\sim N(-1, 1) \\
x_6 - x_{20}   &\sim N(0.5, 1)
\end{align*}
\end{multicols}

\begin{multicols}{3}
\begin{align*}
\bm{\psi}_{1|3}:\\
x_1 &\sim Bern(0.75) \\
x_2 &\sim Bern(0.1) \\
x_3 &\sim Bern(0.3) \\
x_4 &\sim N(0.5, 1.5) \\
x_5 &\sim N(0, 1) \\
x_6 - x_{20} &\sim N(0.5, 1)
\end{align*}

\begin{align*}
\bm{\psi}_{2|3}:\\
x_1 &\sim Bern(0.5) \\
x_2 &\sim Bern(0.3) \\
x_3 &\sim Bern(0.5) \\
x_4 &\sim N(-0.5, 1) \\
x_5 &\sim N(0, 0.5) \\
x_6 - x_{20}   &\sim N(-0.5, 1)
\end{align*}

\begin{align*}
\bm{\psi}_{3|3}:\\
x_1 &\sim Bern(0.5) \\
x_2 &\sim Bern(0.7) \\
x_3 &\sim Bern(0.5) \\
x_4 &\sim N(0, 2) \\
x_5 &\sim N(-1, 2) \\
x_6 - x_{20}  &\sim N(0, 1)
\end{align*}
\end{multicols}

\newpage
\section*{Appendix: Computations}

The R/C++ code is available at on GitHub. We give some further details on updating some of the parameters in our model below.

\textbf{MCMC program:}

\noindent \textbf{Step 0:}
Let $n$ be the total number of subjects and $N$ denote the total number of observations. Initialize all parameter values including $s$, the partitioning variable.
\bigskip

\noindent \textbf{Step 1:} Update $s_i$ for $i = 1, \dots, n$.

Let $n_{\theta}$ be the number of unique $\theta$-clusters in $s_{y,i}$.

\bigskip

\noindent \textbf{Step 2:} Iterate through $k = 1, \dots, n_{\theta}$.

\indent Restrict to subjects with $s_{i, y} = k$. Let $n_k$ be the number of subjects in the cluster and $N_k$ be the total number of observations in this cluster. Below $y_i$, $y$, $X$, $x_i$, $Z$, and $z_i$ will refer to subjects within the given cluster.

\smallskip

\noindent \textbf{Step 2a:} Update $\sigma^{2}_k$ and $\beta^*_k$:

First, calculate residuals: $y_i^* = y_i - z_i \eta_i - u_i$. We specify priors $P(\sigma^2) \sim \text{Inv-Ga}(a_{\beta}, b_{\beta})$ and $P(\bm{\beta}) \sim \text{N}(\bm{\beta_0},\Sigma).$ Define $\Sigma_n = X^{\intercal} X + \Sigma$ and $\beta_n = \Sigma_n^{-1} ( \Sigma \beta_0 + X^{\intercal} y^*)$. The posteriors (within clusters) are given by $P(\sigma^2 | \text{rest}) \sim \text{Inv-Ga}(a_{\beta} + \frac{N_k}{2}, b_{\beta} + \frac{1}{2}(y^{*\intercal}y^* + \beta_0^{\intercal} \Sigma \beta_0 - \beta_n^{\intercal}\Sigma_n \beta_n)$ and $P(\beta | rest) \sim N(\beta_n, \sigma^2_k \Sigma_n)$.

\smallskip

\noindent \textbf{Step 2b:} Update $\sigma^{2}_{b,k}$ and $\eta^*_k$

Now, calculate residuals: $y_i^* = y_i - x_i \beta_i - u_i$. Given prior distributions $P(\sigma_b^2) \sim \text{Inv-Ga}(a_{\beta}, b_{\beta})$ and $P(\bm{\eta}) \sim \text{N}(\mathbf{0},\sigma^2_b \mathbf{I})$, define $\Sigma_{b, n}  = Z^{\intercal} Z / \sigma^2_k + \mathbf{I} / \sigma^2_{b, k}$ and $\mu_n = \left[\sigma^2_{b, n} \right] ^ {-1} Z^{\intercal} y^* / \sigma^2_k$. The posteriors are given by $P(\sigma^2_b | \text{rest}) \sim \text{Inv-Ga}(a_{\beta} + N / 2, b_{\beta} + \frac{1}{2}\bm{\eta}^{\intercal}\bm{\eta})$ and $P(\bm{\eta} | rest) \sim N(\mu_n, [\Sigma_{b, n}]^{-1}$).

\smallskip

\noindent \textbf{Step 2c:} Iterate through $k = 1, \dots, n_{\psi, k}$, where $n_{\psi, k}$ is the number of $\psi$-clusters nested within the $k^{th}$ $\theta$-cluster. Now, we update covariate parameters $\psi$, further restricting to subjects with $s_{i, 2} = k$:

For binary covariates, the prior is $P(p) \sim \text{Beta}(a_x, b_x)$ and the posterior is given by $P(p | \text{rest}) \sim \text{Beta}(\sum_{s = (j, k)} x_{i, l} + a_x, n_{j|k} - \sum_{s = (j, k)} x_{i, l} + b_x).$.

For continuous covariates, prior: $P(\mu, \sigma) \sim \text{scaled Inv-}\chi^2(\nu_0, \tau_0^2) \times \text{N}(\mu_0, \tau^2/c)$ with posteriors $P(\sigma^2 | \text{rest}) \sim \text{scaled Inv-}\chi^2(\nu_0 + n_{\psi, k}, \nu_0 \tau_0 + n_{j|k} * \text{var}(x) + c_0 n_{j|k} / ( c_0 + n_{j|k}) * (\bar{x} - \mu_0)^2)$ and $P(\mu | \text{rest}) \sim N(\mu_n, \sigma^2_n),$ 
where $\sigma^2_n = \frac{1}{c_0 / \tau + n_{j|k} / \tau}$ and $\mu_n = \sigma^2_n ( \mu_0 * c_0 / \tau + \bar{x} n_{j|k} / \tau)$.

This marks the end of the within-cluster updates.

\bigskip

\noindent \textbf{Step 3:} Update random intercept $u_i$:

Iterate through $i = 1, \dots, n$. Calculate residuals $y_i^* = y_i - x_i \beta_i - z_i \eta_i$. Draw $u_i$ from
$N(\mu_u, \sigma^2_{\text{new}})$ where $\sigma^2_{\text{new}} = \frac{\sigma^2_u\sigma^2_i}{n_i \sigma^2_u + \sigma^2_i}$ and $\mu_u = \sigma^2_u \sum_{j = 1}^{n_i} y_{ij} / (n_i \sigma^2_u + \sigma^2_i).$

\noindent \textbf{Step 4:} Update random intercept variance $\sigma^2_u$:

Given prior $P(\sigma^2_u) \sim \text{Inv-Ga}(a_{u}, b_{u})$ the posterior is $P(\sigma^2_u | \text{rest}) \sim \text{Inv-Ga}(a_{u} + \frac{n}{2}, b_{u} + \frac{1}{2} \mathbf{u}^{\intercal} \mathbf{u})$.

\noindent \textbf{Step 5:} Update $\alpha_{\theta}$ (from Escobar and West (1995)):

Let $\alpha_{\theta}$ be the current value. Draw $\gamma \sim Beta(\alpha_{\theta}, n)$. Define $\pi = \frac{n_{\theta} / (n (1 - \log(\gamma)))}{1 + n_{\theta}/(1 - \log(\gamma))}$. Draw $p \sim Bern(\pi)$. Update $\alpha_{\theta}$ from $\text{Gamma}(a_{\alpha} + n_{\theta}, b_{\alpha} - \log(\gamma) )$ with probability $p$ and from $Gamma(a_{\alpha} + n_{\theta} - 1, b_{\alpha} - \log(\gamma) )$ with probability $1 - p$.

\noindent \textbf{Step 6:} Update $\alpha_{\psi}$:

Update $\alpha_{\psi}$ with Metropolis-Hastings step. Our proposal distribution is $\text{Gamma}(a_0, b_0)$. Draw $\alpha_{\text{prop}}$ from proposal distribution. Define
\[
p_1 = \text{dGamma}(\alpha_{\psi}; a_{\alpha}, b_{\alpha}) \alpha_{\psi}^{n_{\theta}} \prod_{j = 1}^{n_{\theta}} \left[(\alpha_{\psi} + n_j) \text{Beta}(\alpha_{\psi} +1, n_j)\right].
\]
Let
\[
p_2 = \text{dGamma}(\alpha_{\text{prop}}; a_{\alpha}, b_{\alpha}) \alpha_{\text{prop}}^{n_{\theta}} \prod_{j = 1}^{n_{\theta}} \left[(\alpha_{\text{prop}} + n_j) \text{Beta}(\alpha_{\text{prop}} +1, n_j)\right].
\]
Note $\text{dGamma}(x; a, b)$ denotes the density of function of a Gamma distribution with parameters $a$ and $b$ evaluated at $x$. $\text{Beta}(u, v)$ denotes the Beta function evaluated at $u$ and $v$. Set $\alpha_{\psi} = \alpha_{\text{prop}}$ with probability $p = \frac{p_2}{p_1}$. Otherwise, use previous $\alpha_{\psi}$.

\bigskip

\noindent Return to Step 1 and repeat until convergence and posteriors are well approximated.

\newpage
\section*{Appendix: Tables - Simulations}

\begin{table}[!ht]
\centering
\begin{tabular} {l c c c c c c}
\hline
& \multicolumn{2}{ c } {EDP} & \multicolumn{2}{ c } {DP} &  \multicolumn{2}{ c } {ME} \\
& $\bar{\ell}_1$ & $\bar{\ell}_2$ & $\bar{\ell}_1$ & $\bar{\ell}_2$ & $\bar{\ell}_1$ & $\bar{\ell}_2$ \\
\hline
$\sigma^2 = 1$; $\sigma^2_u = 0.15$ & 0.66 & 0.87 & 0.91 & 1.46 & 1.10  & 1.83 \\
$\sigma^2 = 1$; $\sigma^2_u = 0.5$ & 0.84 & 1.25 & 1.09 & 1.96 & 1.11 & 1.85 \\
$\sigma^2 = 4$; $\sigma^2_u = 0.15$ & 0.91 & 1.52 & 1.09 & 2.03 & 1.22 & 2.28 \\
$\sigma^2 = 4$; $\sigma^2_u = 0.5$ & 1.06 & 1.92 & 1.18 & 2.33 & 1.23 & 2.33 \\
\hline
\end{tabular}
\caption{Simulation results for $n = 1000$ showing mean $l_1$ and $l_2$ errors over 100 datasets for predictions at $t = 0.75$. $\sigma^2$ indicates the simulated regression variance and $\sigma^2_u$ indicates the simulated random intercept variance. EDP indicates the longitudinal model with an enriched Dirichlet process prior. DP indicates the longitudinal model with a Dirichlet process prior. ME indicates a mixed effects model fit using the lmer package in R. Fit with cubic B-splines with 2 knots.}
\end{table}

\begin{table}[!ht]
\centering
\begin{tabular} {l c c c c c c}
\hline
& \multicolumn{2}{ c } {EDP} & \multicolumn{2}{ c } {DP} &  \multicolumn{2}{ c } {ME} \\
& $\bar{\ell}_1$ & $\bar{\ell}_2$ & $\bar{\ell}_1$ & $\bar{\ell}_2$ & $\bar{\ell}_1$ & $\bar{\ell}_2$ \\
\hline
$\sigma^2 = 1$; $\sigma^2_u = 0.15$ & 0.62 & 0.76 & 0.92 & 1.49 & 1.10 & 1.81 \\
$\sigma^2 = 1$; $\sigma^2_u = 0.5$ & 0.77 & 1.07 & 1.10 & 2.02 & 1.11 & 1.85 \\
$\sigma^2 = 4$; $\sigma^2_u = 0.15$ & 0.72 & 1.03 & 1.04 & 1.90 & 1.21 & 2.26 \\
$\sigma^2 = 4$; $\sigma^2_u = 0.5$ & 0.85 & 1.28 &1.15 & 2.26 & 1.23 & 2.33  \\
\hline
\end{tabular}
\caption{Simulation results for $n = 5000$ showing mean $l_1$ and $l_2$ errors over 100 datasets for predictions at $t = 0.75$. $\sigma^2$ indicates the simulated regression variance and $\sigma^2_u$ indicates the simulated random intercept variance. EDP indicates the longitudinal model with an enriched Dirichlet process prior. DP indicates the longitudinal model with a Dirichlet process prior. ME indicates a mixed effects model fit using the lmer package in R. Fit with cubic B-splines with 2 knots.} 
\end{table}


\begin{table}[!ht]
\centering
\begin{tabular} {l c c c c c c}
\hline
& \multicolumn{2}{ c } {EDP} & \multicolumn{2}{ c } {DP} &  \multicolumn{2}{ c } {ME} \\
& $\bar{\ell}_1$ & $\bar{\ell}_2$ & $\bar{\ell}_1$ & $\bar{\ell}_2$ & $\bar{\ell}_1$ & $\bar{\ell}_2$ \\
\hline
$\sigma^2 = 1$; $\sigma^2_u = 0.15$ & 0.31 & 0.15 & 0.31 & 0.15 & 0.26 & 0.11 \\
$\sigma^2 = 1$; $\sigma^2_u = 0.5$ & 0.56 & 0.50 & 0.56 & 0.50 & 0.37 & 0.22 \\
$\sigma^2 = 4$; $\sigma^2_u = 0.15$ & 0.32 & 0.16 & 0.32 & 0.16 & 0.31 & 0.15 \\
$\sigma^2 = 4$; $\sigma^2_u = 0.5$ & 0.57 & 0.50 &0.57 & 0.50 & 0.49 & 0.38  \\
\hline
\end{tabular}
\caption{Simulation results for $n = 5000$ showing mean $l_1$ and $l_2$ errors over 100 datasets for predictions at $t = 0.75$. $\sigma^2$ indicates the simulated regression variance and $\sigma^2_u$ indicates the simulated random intercept variance. EDP indicates the longitudinal model with an enriched Dirichlet process prior. DP indicates the longitudinal model with a Dirichlet process prior. ME indicates a mixed effects model fit using the lmer package in R. Fit with cubic B-splines with 2 knots.} 
\end{table}

\newpage
\section*{Appendix: R code for choosing number of clusters}

\begin{verbatim}
## function for calculating n x n matrix of how many times 
## subjects in same cluster
adjmatrix <- function(s) {
  n <- nrow(s[[1]])
  mat <- matrix(0, n, n)
  nelem <- length(s)
  for(i in 1:nelem){
    temp.mat <- as.integer(outer( s[[i]][ ,1], s[[i]][ ,1], FUN = "==" ) )
    mat <- mat + temp.mat
  }
  return(mat)
}

## a1c.res$s is a list of cluster memberships 
## for successive MCMC iterations
## each element is a n x 2 matrix 
## the first column is the theta-cluster membership
## the second column is the psi-cluster subcluster membership
a1c.adj <- adjmatrix(a1c.res$s)
a1c.dist <- dist(a1c.adj, method = "maximum")
hi <- hclust(a1c.dist, method = "ward.D2")
clust.a1c <-cutree(hi, k = 2)
\end{verbatim}

\newpage

\section*{Appendix: Various computations}

Discrete covariates:
\begin{align*}
\int p^x (1 - p)^{1-x} \frac{p^{\alpha - 1} (1-p)^{\beta - 1}}{\text{Be}(\alpha, \beta)}dp & = \frac{1}{\text{Be}(\alpha, \beta)} \int p^{\alpha + x -1}(1 - p)^{\beta - x} dp\\
& = \frac{\text{Be}(\alpha + x, \beta - x + 1)}{\text{Be}(\alpha, \beta)}
\end{align*}

Continuous covariates:\\
Prior: Normal-inverse-$\chi$-squared:

\begin{align*}
p(\mu, \sigma^2) &= \frac{\sqrt{c_0}}{\sqrt{2\pi}\sqrt{\tau_0}} \exp\left(\frac{-(\mu-\mu_0)^2}{2\tau_0/c_0}\right)\frac{(\tau_0\nu_0/2)^{\nu_0/2}}{\Gamma(\nu_0/2)} \frac{\exp\left(\frac{-\nu_0\tau_0}{2\sigma^2}\right)}{(\sigma^2)^{1+\nu_0/2}} \\
& =  \frac{\sqrt{c_0}}{\sqrt{2\pi}\sqrt{\tau_0}} \frac{(\tau_0\nu_0/2)^{\nu_0/2}}{\Gamma(\nu_0/2)} \exp\left(\frac{-(\mu-\mu_0)^2}{2\tau_0/c_0}\right) \frac{\exp\left(\frac{-\nu_0\tau_0}{2\sigma^2}\right)}{(\sigma^2)^{1+\nu_0/2}} \\
& \propto \exp\left(\frac{-(\mu-\mu_0)^2}{2\tau_0/c_0}\right) \frac{\exp\left(\frac{-\nu_0\tau_0}{2\sigma^2}\right)}{(\sigma^2)^{1+\nu_0/2}}
\end{align*}

Data:

\[
p(x | \mu, \sigma^2) =  \frac{1}{\sqrt{2\pi}\sqrt{\sigma^2}}\exp\left(\frac{-(x - \mu)^2}{2\sigma^2}\right)
\]

Posterior:

\[
p(\mu, \sigma^2 | x) \propto \sigma^{-3} (\sigma^2)^{-(\nu_n/2)} \exp\left(-\frac{1}{2\sigma^2}[\nu_n\sigma^2_n + c_n(\mu_n - \mu)^2]\right)
\]

\begin{align*}
h(x) &= \int \int p(\mu, \sigma^2 | x) d \mu \ d \sigma^2 \\
&= \frac{1}{\sqrt{2\pi}} \frac{\sqrt{c_0}}{\sqrt{2\pi}\sqrt{\tau_0}} \frac{(\tau_0\nu_0/2)^{\nu_0/2}}{\Gamma(\nu_0/2)} \frac{\sqrt{2\pi}\sqrt{\tau_n}}{\sqrt{c_n}} \frac{\Gamma(\nu_n/2)} {(\tau_n\nu_n/2)^{\nu_n/2}} \\
&=  \frac{1}{\sqrt{2\pi}} \frac{c_0}{c_n} \frac{\tau_n}{\tau_0} \frac{(\tau_0\nu_0/2)^{\nu_0/2}}{(\tau_n\nu_n/2)^{\nu_n/2}}\frac{\Gamma(\nu_n/2)}{\Gamma(\nu_0/2)}
\end{align*}
where
\begin{align*}
c_n &= c_0 + 1 \\
\nu_n &= \nu_0 + 1\\
\tau_n &= \frac{1}{\nu_n}\left(\nu_0\tau_0 + \frac{c_0}{c_n}\left(\mu_0 - x\right)^2\right)
\end{align*}

\end{document}